\documentclass[NewProceedings]{ascelike}
\usepackage{amsmath}
\usepackage{graphicx,caption,subcaption}
\usepackage{bm}
\usepackage{graphicx}

\usepackage{url}

\usepackage{authblk}
\usepackage{subcaption}
\usepackage{float}
\usepackage{url,hyperref,lineno,microtype}
\usepackage{graphics,graphicx,color}
\usepackage{caption}
\usepackage[ruled,longend]{algorithm2e}
\usepackage{amsmath,amssymb,amsthm}
\usepackage{longtable,float}
\usepackage{epsfig}
\usepackage[utf8]{inputenc}
\usepackage[english]{babel}
\usepackage{comment}
\usepackage{multirow}
\usepackage[none]{hyphenat}
\usepackage{indentfirst}

\renewcommand{\vec}[1]{\ensuremath{#1}}
\newcommand{\T}{\ensuremath{^\mathsf{T}}}
\graphicspath{ {./figures/} }
\DeclareGraphicsExtensions{.png}
\usepackage{lipsum}
\makeatletter
\renewcommand{\maketitle}{\bgroup\setlength{\parindent}{0pt}
  \textbf{\@title}
\begin{flushleft}
  \@author
\end{flushleft}\egroup
}
\renewcommand{\vec}[1]{\ensuremath{#1}}

\usepackage{listings}
\usepackage{booktabs} 
\usepackage{scalerel}

\newcommand\reallywidehat[1]{\arraycolsep=0pt\relax%
\begin{array}{c}
\stretchto{
  \scaleto{
    \scalerel*[\widthof{\ensuremath{#1}}]{\kern-.5pt\bigwedge\kern-.5pt}
    {\rule[-\textheight/2]{1ex}{\textheight}} %WIDTH-LIMITED BIG WEDGE
  }{\textheight} % 
}{0.5ex}\\           % THIS SQUEEZES THE WEDGE TO 0.5ex HEIGHT
#1\\                 % THIS STACKS THE WEDGE ATOP THE ARGUMENT
\rule{-1ex}{0ex}
\end{array}
}
\makeatother

\title{\begin{center} Deployable Tensegrity Lunar Tower \end{center}}
\author[1]{Muhao Chen}
\author[1]{Raman Goyal}
\author[2]{Manoranjan Majji}
\author[3]{Robert E. Skelton}
\affil[1]{\noindent Graduate Student, Department of Aerospace Engineering, Texas A\&M University, TX, USA; emails: muhaochen@tamu.edu, ramaniitrgoyal92@tamu.edu}
\affil[2]{\noindent Assistant Professor, Department of Aerospace Engineering, Texas A\&M University, TX, USA; email: mmajji@tamu.edu}
\affil[3]{\noindent TEES Eminent Research Professor, Department of Aerospace Engineering, Texas A\&M University, TX, USA; email: bobskelton@tamu.edu}

\begin{document}
\maketitle
% Please include an abstract:
\section*{ABSTRACT}

\noindent A tensegrity tower design to support a given payload for the moon mining operation is proposed in this paper. A non-linear optimization problem for the minimal-mass structure design is posed and solved, subject to the yielding constraints for strings and yielding and buckling constraints for bars in the presence of lunar gravity. The optimization variables for this non-linear problem are structural complexity and pre-stress in the strings. Apart from local failure constraints of yielding and buckling, global buckling is also considered. The structure designed as a deployable tower is a $T_nD_1$ tensegrity structure. A case study demonstrates the feasibility and advantage of the tower design. The principles developed in this paper are also applicable for building other structures on the Earth or other planets.

\section{Introduction}
~

\noindent 
%After about 60 years of space exploration, it has been shared in some degree the belief – made concrete by the Apollo program, but shared with all who ever looked up at the night sky and wondered - that mankind will leave the earth cradle and destine to stars. 
\noindent The great heroic success of the Apollo program has triggered a strong will, passion, and enthusiasm of the public. Till now, humans have developed various heavy rockets, put rovers on other planets, sent people to ISS, and launched probes to the Sun. The interest is now returning to the moon to utilize its abundant resources. Various Lunar exploration missions have provided us with information about its abundant useful resources. For example, the entire lunar surface is covered with an unconsolidated layer of regolith \cite{heiken1991lunar}, which can be used as a very efficient material for building a space habitat shield \cite{chen2018energy}. The moon is especially rich in Ca, Al, Si, O, Mg, Fe, and Ti \cite{crawford2015lunar}. The recent study also shows that a large amount of ice has a permanent presence in the shadowed lunar polar craters \cite{spudis2013evidence}. The goal of this paper is to study a feasible design of a lightweight tower to support moon mining operations. \\

\noindent Scientists have been exploring the idea of mining the moon for some time. Mining ice on the moon can be achieved by using solar energy to heat the ice and store it as water \cite{duke1998mining}. Rock breakage by microwave techniques, mineral processing, and materials manufacturing for ISRU have also been discussed \cite{tukkaraja2018lunar}. Tunnel Boring Machines could offer another safe and efficient approach for mining on the moon \cite{rostami2018lunar}. Sanders presented NASA's lunar ISRU strategy, which includes plans for regolith, polar water/volatile mining, commercial opportunities, rovers, and mission schedules \cite{sanders2019nasa}. The mining design of these works mainly focuses on rovers, operations, and extracting minerals. Some of the important issues are left unsolved for mining in the permanent shaded polar craters at the high latitude of the moon: 1) a structure to help collect and distribute solar energy efficiently (mirrors and solar panels to light up the operation area, store energy, and generate heat) and 2) supporting communication equipment. All these problems lead to the requirement of a lunar tower. As mass is one of the most critical issues for space exploration, we desire to design a minimal mass deployable lunar tower. \\

\noindent Tensegrity system is a subset of multi-body systems, which includes cylindrical rigid bodies (bars) and elastic members (strings) arranged in a stabilizable topology \cite{Skelton_2009_Tensegrity_Book}. The tensegrity art-form was first created by Ioganson (1921) and Snelson (1948), and the word \lq Tensegrity\rq ~was coined as \lq Tensile + Integrity = Tensegrity\rq~by Buckminster Fuller \cite{Fuller_1959}. After decades of study, tensegrity structures have shown their great advantage in designing lightweight structures  \cite{Skelton_2009_Tensegrity_Book}. Tensegrity system, as a new dimension of engineering thought, motivates engineers to rethink and study structures in a more fundamental way, for example, the mechanical response of 3D tensegrity lattices \cite{rimoli2017mechanical}, minimum mass bridges \cite{fabbrocino2017optimal}, high-performance robotics \cite{bliss2012experimental}, lander \cite{sunspiral2013tensegrity,chen2017soft,Goyal_2018_Buckling,zhao2019theoretical}, tensegrity spine \cite{sabelhaus2017model}, and perhaps biology shows the great evidence that tensegrity should be the way to design structures \cite{Skelton_2009_Tensegrity_Book}. Sultan and Skelton gave a deployable tensegrity tower based on a multi-stage three-strut Snelson-type tensegrity topology \cite{sultan2003deployment}. Schlaich built probably the tallest tower (62.3 m, consists of 6 three-strut Snelson-type tensegrity unit, each unit 8.3 m) \cite{schlaich2004messeturm}. Yildiz and Lesieutre studied stiffness properties and deployable strategies of a class-1 and class-2 Snelson-type tower \cite{yildiz2018novel,yildiz2019effective}. However, none of these towers gives a complete description of the minimal mass deployable tower design in the presence of gravity. This paper utilizes the tensegrity paradigm to design a tower for lunar mining, requiring less mass to meet load constraints. The principles developed in this paper can also be used for other tensegrity structure designs.\\

\noindent This paper is organized as follows: Section 1 provides motivation and introduction of the lunar tower design for mining on the moon. Section 2 derives the minimum mass tensegrity principles, including statics analysis, structure mass formulation, gravitational forces, and global stability analysis. To solve the nonlinear optimization problem, an iterative algorithm is also provided. Section 3 describes the deployable $T_nD_1$ tower design, which is a combination of T-Bar and D-Bar systems, and Section 4 presents a case study detailing the results of minimum mass design. Section 5 provides a concise summary of the research work.
~

\section{Minimal Mass Tensegrity Principles}
~

\subsection{Static Analysis for Tensegrity Structures}
~

\noindent The static equilibrium equation for a given tensegrity structure and given external force can be written as \cite{Goyal_Dynamics_2019}:
\begin{equation}\label{statics}
  NK = W, ~ K = C_s\T \hat{\gamma}C_s - C_b\T \hat{\lambda} C_b,
\end{equation}
where $N\in R^{3 \times n}$ is the nodal matrix with each column of $N$ representing the node position, $n$ is number of nodes, and, $C_s\in R^{\alpha \times n}$ and $C_b \in R^{\beta \times n}$ are the connectivity matrices of strings and bars (with 0, -1, and 1 contained in each row), respectively. The number of bars and strings are denoted by $\alpha$ and $\beta$. The external force matrix $W\in R^{3\times n}$ contains each of its column as the force vector acting on the corresponding node, $\gamma \in R^{\alpha \times 1}$ is a vector of force densities (force per unit length) in the strings, and $\hat{\vec{v}}$ is a diagonal matrix of the elements of a vector \vec{v}. The string and bar vectors are contained in the string matrix $S= \begin{bmatrix} s_1  & s_2 & \cdots & s_{\alpha} \end{bmatrix}\in R^{3\times \alpha}$ and in the bar matrix $B= \begin{bmatrix} b_1 & b_2 & \cdots & b_{\beta} \end{bmatrix} \in R^{3\times \beta}$ respectively, such as $S = NC_s^T$ and $B = NC_b^T$.  Let us take the $i$\textsuperscript{th} column of Eq.~(\ref{statics}):
\begin{align}
    S\hat{\gamma}C_s e_i-B \hat{\lambda} C_b e_i=We_i,
\end{align}
where $e_i = [0 ~ 0 ~ \cdots ~ 1 ~ \cdots ~ 0 ~ 0]^T$ is a column with 1 as the $i$\textsuperscript{th} element and zeros elsewhere. Using the identity $\hat{x}y=\hat{y}x$, for $x$ and $y$ being column vectors, we can write the previous equation as:
\begin{align}
 S\reallywidehat{(C_s e_i)}\gamma-B \reallywidehat{( C_b e_i)}\lambda= W{e_i}.
\end{align}\par

\noindent Stacking all the columns from $i = 1$ to $i = n$\textsuperscript{th} column, we get:
\begin{align}\label{equality}
Ax & = W_{vec}, ~x = \begin{bmatrix} \gamma^T & \lambda^T \end{bmatrix}^T,
\end{align}
where:
\begin{align}
    A = \begin{bmatrix}
    S\reallywidehat{(C_s e_1)}&-B \reallywidehat{( C_b e_1)} \\
    S\reallywidehat{(C_s e_2)}&-B \reallywidehat{( C_b e_2)} \\
    \vdots& \vdots \\
    S\reallywidehat{(C_s e_n)}&-B \reallywidehat{( C_b e_n)}
    \end{bmatrix}, ~W_{vec} = \begin{bmatrix}
    We_1 \\ We_2 \\ \vdots \\ We_n 
    \end{bmatrix}.
\end{align}
~

%  \\
% (A^T)_i & = S\reallywidehat{(C_s e_i)}-B \reallywidehat{( C_b e_i)}~\text{and}~ W_{{vec}~i} = W{e_i}.

\subsection{Mass Formulation for Tensegrity Structures}
~

\noindent The strings can fail by material yielding while bars can fail by both yielding and buckling. The minimum mass of the structure is obtained when all the strings are designed to their yield point and all the bars are designed at the onset of yielding or buckling:

\begin{equation}\label{objective}
M =  \frac{\rho_s}{\sigma_s} \sum_{i = 1}^{\alpha} {\gamma_i}||s_i||^2 + \sum_{j = 1}^{\beta} \max{\left( \frac{\rho_b}{\sigma_b}{\lambda_j}||b_j||^2, 2\rho_b \lambda_j^{\frac{1}{2}}\left(\frac{||b_j||^5}{\pi E_b}\right)^{\frac{1}{2}}\right)},
\end{equation}
where $\rho_s$, $\rho_b$, $\sigma_s$, $\sigma_b$ are the density and yield strength of strings and bars, respectively.
The length of each string and each bar are denoted by $||s_i||$, and $||b_j||$ for $i = 1, 2, \cdots, \alpha$; and  $j = 1, 2, \cdots, \beta$, and $E_b$ is Young's modulus of bars. As seen from Eq.~(\ref{objective}), the maximum of mass required for yielding and buckling is considered for each bar.\\
~
\begin{comment}
The criteria for yielding and buckling of each bar can be written as 
\begin{equation}
    \frac{m_Y}{m_B} = \frac{1}{2\sigma_b}\sqrt{\frac{\lambda_j \pi E_b}{||b_j||}}, ~ j = 1, 2, \cdots, \beta.
\end{equation}
For a given $\lambda_j$ and $||b_j||$, a bar fails due to yielding if $m_Y \textgreater m_B$, a bar fails due to buckling if $m_Y\textless m_B$, a bar failure due to buckling and yielding happens at the same time if $m_Y = m_B$. Or discuss focusing on the force densities of bars: 
a bar fails under buckling if $\lambda \textless \frac{4\sigma_b^2||b||}{\pi E_b}$, a bar fails under yielding if $\lambda \textgreater \frac{4\sigma_b^2||b||}{\pi E_b}$, failure of a bar under yielding and buckling happens at the same time if $\lambda = \frac{4\sigma_b^2||b||}{\pi E_b}$.\par
\end{comment}

\noindent Let us define a label matrix $Q\in R^{\beta \times \beta}$ as:
\begin{equation}\label{solid_q_matrix}
Q_{jj} = \left \{
\begin{aligned}
& 0; & & \lambda_j \geq \frac{4\sigma_b^2||b_j||}{\pi E_b},~ \text{bar yields} \\
& 1; & & \lambda_j ~ \textless ~ \frac{4\sigma_b^2||b_j||}{\pi E_b},~ \text{bar buckles}
\end{aligned}
\right.,
\end{equation}
and the off diagonal elements of $Q$ are zeros. This matrix is used to identify whether a bar is yielding or buckling. Notice that a simple operation of $(I-Q)$ works intuitively for representing the bar under yielding constraint. The bars are now separated into two types and the minimal mass formula can be well defined in matrix form \cite{Goyal_MOTES_2019}:  
\begin{multline}\label{yb_case}
    M =  \frac{\rho_s}{\sigma_s}(vec{(\lfloor S^TS \rfloor))^T}\gamma + \frac{\rho_b}{\sigma_b}(vec{(\lfloor B^TB \rfloor(I-Q)))^T}\lambda \\+ \frac{2\rho_b}{\sqrt{\pi E_b}} (vec{(\lfloor B^TB\rfloor ^{\frac{5}{4}}Q)})^T \lambda^{\frac{1}{2}},
\end{multline}
where $\lfloor \bullet \rfloor$ is an operator taking the diagonal elements of a matrix, $vec(\bullet)$ is an operator taking the elements of the matrix and form a vector. \\
~

\subsection{Tensegrity Statics in the Presence of Gravity}
~

\noindent Since gravity is unlike a given set of specific external forces that apply to the structure, it is determined by the mass of the structure itself. In other words, the statics mass optimization process is coupled with the gravity force. The total force $W$ can be separated into two parts $W = W_e + W_g$, where $W_g$ is the gravity force, and $W_e$ is other applied external force. The acceleration due to gravity is defined as $\bm{g}$ with lunar gravity as  $\bm{g}_{moon} = \left[ \begin{matrix} 0& 0& -1.62 \end{matrix}\right]^T$ m/$s^2$. The gravity force can be modeled by lumped forces equally distributed on the member nodes \cite{nagase2014minimal}. Thus, the gravitational force due to bars and strings can be expressed as:
\begin{multline}
     W_{gi}  = \frac{1}{2}\bm{g} \frac{\rho_s}{\sigma_s}\left(vec(\lfloor S^TS \rfloor)\right)^T\reallywidehat{|C_s e_i|}\gamma  + \frac{1}{2}\bm{g}\frac{\rho_b}{\sigma_b}\left(vec(\lfloor B^TB \rfloor(I-Q))\right)^T \reallywidehat{|C_b e_i|}\lambda \\ 
     + \frac{1}{2}\bm{g}\frac{2\rho_b}{\sqrt{\pi E_b}} \left(vec{(\lfloor B^TB\rfloor ^{\frac{5}{4}}Q)}\right)^T \reallywidehat{|C_b e_i|}\lambda^{\frac{1}{2}},\label{gravity_solid_bar}
\end{multline}
where $|\bullet|$ is an operator getting the absolute value of each element for a given matrix. Stacking all the columns, we get:
 \begin{align}\label{gravity_vec}
    W_{g~vec} &=  \begin{bmatrix}  W_{g1}^T &  \cdots &  W_{gi}^T & \cdots &  W_{gn}^T \end{bmatrix}^T.
\end{align}\\
~

\subsection{Global Stiffness Matrix for Tensegrity Structures}
~

\begin{sloppypar}
\noindent Consider a small variation around the equilibrium position as:
\begin{align}
    N+dN = \begin{bmatrix} \cdots & (n_k+dn_k) & \cdots \end{bmatrix}, ~ W+dW = \begin{bmatrix} \cdots & (w_k+dw_k) & \cdots \end{bmatrix}, \\
    \gamma+d\gamma = \begin{bmatrix} \cdots & (\gamma_i+d\gamma_i) & \cdots \end{bmatrix}^T, ~ \lambda+d\lambda = \begin{bmatrix} \cdots & (\lambda_j+d\lambda_j) & \cdots \end{bmatrix}^T,
\end{align}
and write the statics equation about the equilibrium as \cite{nagase2014minimal}:
\begin{align}\label{variation}
    (N+dN)C_s^T (\reallywidehat{\gamma+d\gamma}) C_s - (N+dN) C_b^T (\reallywidehat{\lambda+d\lambda}) C_b = W+dW.
    \end{align}\par
\noindent We assume the materials are Hookean and therefore the force densities (string in tension, bar in compression) can be expressed as:
\begin{align}\label{force_den}
    \gamma_{i} = k_{si} \left(1-\frac{||s_{i0}||}{||s_i||}\right), ~\lambda_{j} = -k_{bj}\left(1-\frac{||b_{j0}||}{||b_j||}\right),
\end{align}
where $||s_{i0}||$ and $||b_{j0}||$ are the rest length of the $i^{th}$ string and $j^{th}$ bar. The spring constants of strings and bars, $k_{si}$ and $k_{bj}$ satisfy:
\begin{align}\label{spring}
    k_{si}=\frac{E_{si}A_{si}}{||s_{i0}||}, ~ k_{bj}=\frac{E_{bj}A_{bj}}{||b_{j0}||},
\end{align}
where $A_{si}$ and $A_{bj}$ are cross-section areas,  $E_{si}$ and $E_{bj}$ are Young's modules of the strings and bars. With the information of label matrix $Q$, mass of a string and a bar are given as:
\begin{align}
\label{mass_memebr}    m_{si} = \frac{\rho_s}{\sigma_s}||s_i||^2\gamma_i, ~ m_{bj} = (1-Q_{jj})\frac{\rho_b}{\sigma_b}||b_j||^2\lambda_j+Q_{jj}\frac{2\rho_b}{\sqrt{\pi E_{bj}}}||b_j||^{\frac{5}{2}}\lambda_j^{\frac{1}{2}}.
\end{align}\par
\end{sloppypar}

\noindent Rearrange Eqs.~(\ref{variation}-\ref{mass_memebr}), we get the stiffness matrix $K_n$ for tensegrity structure subject to yielding and buckling constraints as:
\begin{align}
    K_n vec(dN) = vec(dW),
\end{align}
where
\begin{multline}\label{Kn}
K_n = (C_s^T \otimes I_3)\textbf{b.d.($K_{s1}$,$\cdots$,$K_{s\alpha}$})(C_s \otimes I_3) \\-  
(C_b^T \otimes I_3)\textbf{b.d.($K_{b1}$, $\cdots$, $K_{b\beta}$})(C_b \otimes I_3),
\end{multline}
and \begin{align}
\nonumber
    K_{si} & = \gamma_i \left(I_3 + \frac{E_{si}}{\sigma_s}\frac{s_i s_i^T}{||s_i||^2}\right) ,\\
       K_{bj} & = \lambda_j \left(I_3 - (1-Q_{jj})\frac{E_{bj}}{\sigma_b}\frac{b_j b_j^T}{||b_j||^2}\right) - 2Q_{jj} \sqrt{\frac{E_{bj}}{\pi}}\frac{b_j b_j^T}{||b_j||^{\frac{3}{2}}}\lambda_j^{\frac{1}{2}}.
\end{align}\par
~
\subsection{Algorithm for Minimal Mass Tensegrity Structure}
~

\noindent The minimal mass problems can be formulated as:
\begin{equation}\label{solid_yb_min}
\left \{
\begin{aligned}
& \underset{x}{\text{minimize}}
& & M \\
& \text{subject to}
& & Ax = W_{e~vec} + W_{g~vec}, ~ x \geq \epsilon_0, \text{and}~ eig(K_n) > \mu I
\end{aligned}
\right.,
\end{equation}
where $\epsilon_0$ is the prestress assigned to the strings, and $\epsilon_0 \geq 0$ guarantees that all strings are in tension and all bars in compression, $eig(K_n)$ returns the eigenvalues of the matrix $K_n$, and the system is globally stable at the equilibrium for $\mu \geq 0$.

\noindent Notice that to solve Eq.~(\ref{solid_yb_min}), one needs to specify the label matrix $Q$. However, one cannot exactly tell $Q$ for any structure in advance because it is determined by structure topology and external force. To obtain a global solution, the nonlinear optimization problem can be solved in an iterative manner as described in Algorithm~\ref{alg1}.\par
~

\begin{algorithm}[h!]
\caption{Minimal Mass Tensegrity subject to  Stability and Gravity}
{\bf 1)}  Let $Q={I}^{\beta \times \beta}$, $W_{g~vec} = 0$, $\epsilon_0 = 0$, $\delta \epsilon = 0.01 $, $\mu = 0$.\\
{\bf 2)} Compute force densities $x$: \\
\While{$\min\{eig(K_n)\}<\mu$}{
\While{$Q_{i+1} \neq Q_{i}$}{
\begin{align}\nonumber 
&\begin{cases}
\underset{x}{\text{minimize}}~
M \\ \text{subject to}~
Ax = W_{e~vec} + W_{g~vec}, ~ x \geq \epsilon_0. 
\end{cases} \\ \nonumber 
& \text{Compute $\lambda$ from $x$, check Eq.~(\ref{solid_q_matrix}), update $Q$.}\\ \nonumber
& \text{Update $W_{g~vec}$ from Eq.~(\ref{gravity_solid_bar})}.
\\ \nonumber
& i \leftarrow i+1.
\end{align}
}
Compute stiffness matrix $K_n$ from Eq.~(\ref{Kn}).\\
$\epsilon_0 \leftarrow \epsilon_0 + \delta \epsilon$.}.
\label{alg1}
\end{algorithm}

\section{Deployable Tower Design}
~
\subsection{T-Bar Structure}
~

\begin{figure}[h!]
\begin{subfigure}{0.5\textwidth}
    \centering
    \includegraphics[width=1\textwidth]{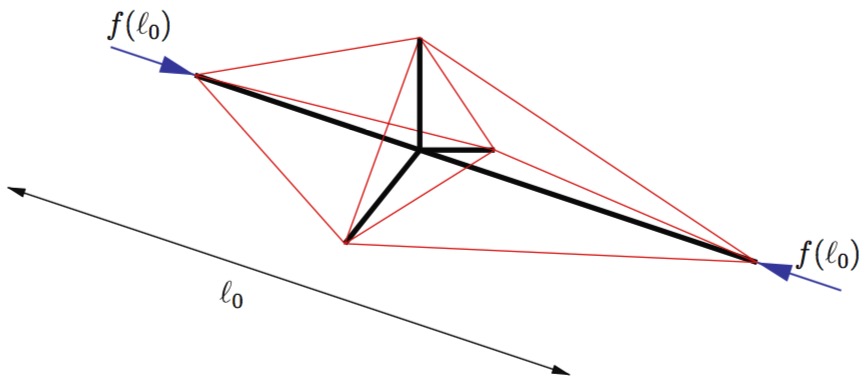}
    \caption{3D T-Bar unit.}
     \label{3DTbar}
\end{subfigure}
\begin{subfigure}{0.5\textwidth}
    \centering
    \includegraphics[width=1\textwidth]{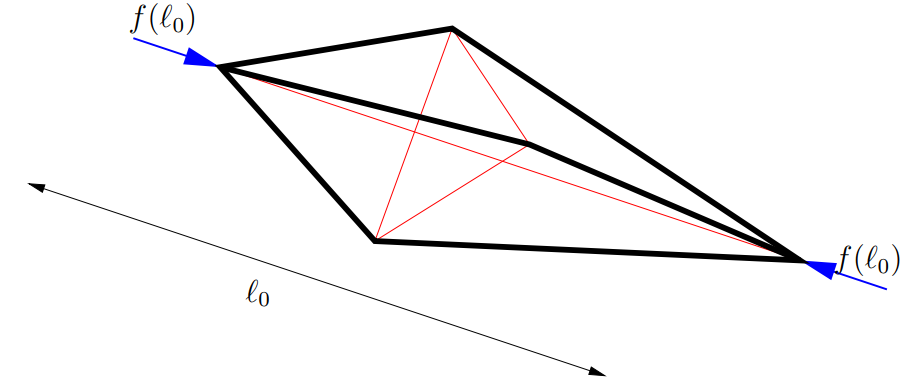}
    \caption{3D D-Bar unit.}
    \label{3DDBar}
\end{subfigure}
\caption{Three-dimensional tensegrity T-Bar and D-Bar unit, black lines are bars and red lines are strings.}
% \label{DHT_habitat}
\end{figure}

\begin{figure}[h!]
\begin{subfigure}{0.5\textwidth}
    \centering
    \includegraphics[width=1\textwidth]{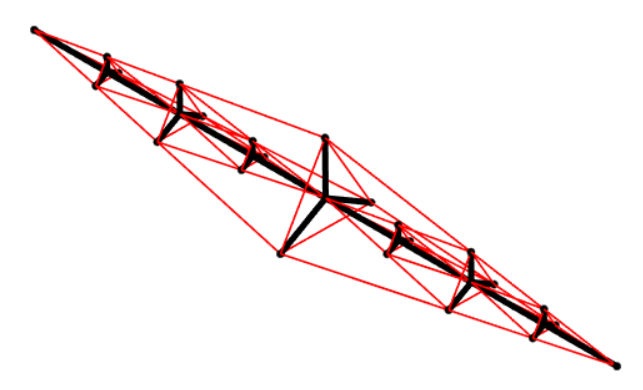}
    \caption{Three-dimensional $T_3$ structure.}
     \label{3d_tbar_system}
\end{subfigure}
\begin{subfigure}{0.5\textwidth}
    \centering
    \includegraphics[width=0.9\textwidth]{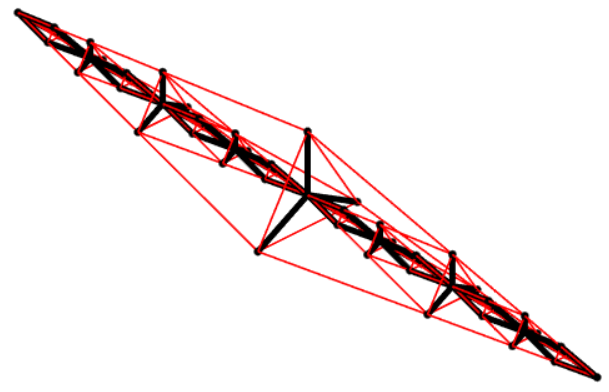}
    \caption{Three-dimensional $T_3D_1$ structure.}
    \label{3d_tdbar_system}
\end{subfigure}
\caption{Three-dimensional tensegrity T-Bar and $T_nD_1$-Bar structure.}
% \label{DHT_habitat}
\end{figure}

\noindent Skelton and Oliveira have proved that T-Bar and D-Bar systems require less mass than a continuum bar in taking the same compression load $f(l_0)$. A three-dimensional T-Bar unit is shown in Figure~\ref{3DTbar} \cite{Skelton_2009_Tensegrity_Book}. Each longitudinal bar in the T-bar structure can be replaced with another T-bar tensegrity unit while preserving the total length of the structure. Repeating this self-similar process $q$ times is defined as the complexity of the structure. Figure~\ref{3d_tbar_system} shows a T-Bar structure of complexity $q=3$.\\
% There exists an optimal complexity for a given compressive load $f(l_0)$ \cite{Skelton_2009_Tensegrity_Book}. \\
~

\subsection{D-Bar Structure}
~

\noindent The dual of the T-Bar unit is called a D-Bar structure, which is shown in Figure~\ref{3DDBar}.
The D-bar is also shown to be a more efficient structure in taking a compressive load than a continuum bar \cite{Skelton_2009_Tensegrity_Book}. Another advantage of the D-bar structure is its deployability. The length of the structure can be changed by controlling the length of the individual strings. \\
~

\subsection{Tensegrity Tower Design}
~

\begin{figure}[h!]
\begin{subfigure}{0.5\textwidth}
    \centering
    \includegraphics[width=0.7\textwidth]{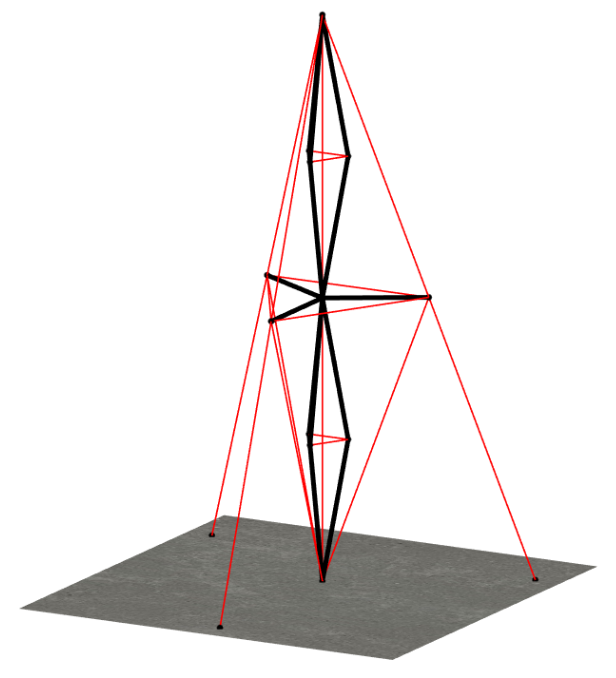}
    \caption{Mathematical tower model.}
     \label{math_model}
\end{subfigure}
\begin{subfigure}{0.5\textwidth}
    \centering
    \includegraphics[width=0.6\textwidth]{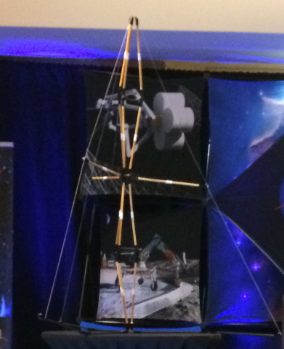}
    \caption{Experimental tower model.}
    \label{exper_model}
\end{subfigure}
\caption{Three-dimensional $T_2D_1$ tensegrity tower models, source of (b): \url{http://www.leonarddavid.com/lunar-polar-propellant-mining-outpost-envisioned/}.}
\label{models}
\end{figure}

\noindent We combine the two structures such that it is both deployable and mass efficient in taking compression.
Let us start with a T-Bar structure with complexity $q = n$ and replace its each $2^n$ longitudinal bars with D-Bar units of complexity 1 to obtain a $T_nD_1$ structure. A 3-dimensional $T_3 D_1$ is shown in Figure~\ref{3d_tdbar_system}. The same $T_n D_1$ structure is used to design towers as the payload on top of the tower will exert a compressive load on two ends of the structure. The simulation and experimental model is shown in Figure~\ref{math_model}.

\subsection{Tower Deployment Discussion}
~

\begin{figure}[h!]
    \centering
    \includegraphics[width=1\textwidth]{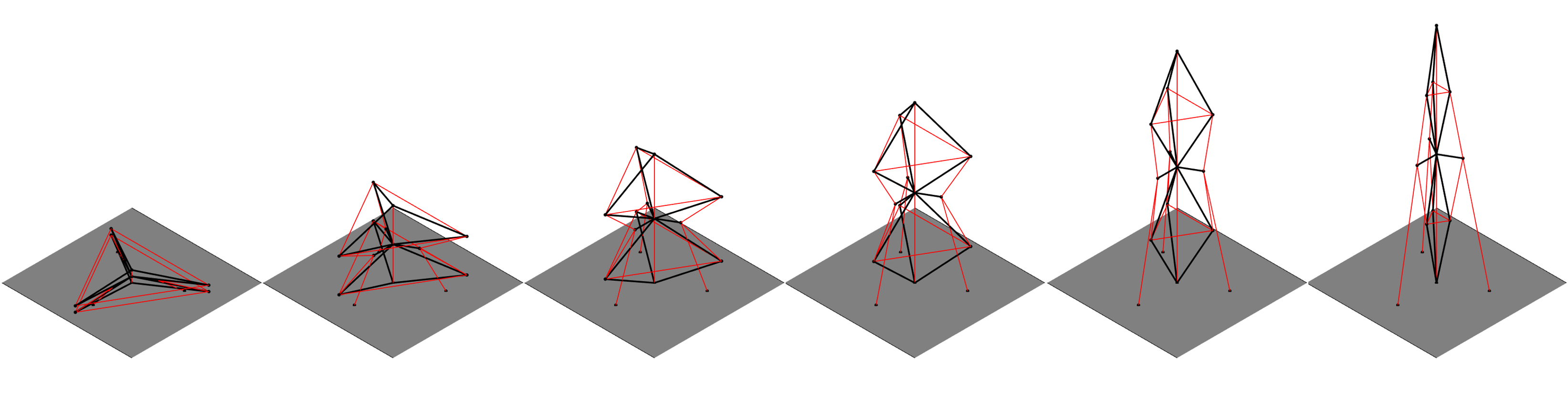}
    \caption{Tower deployment from a stowed configuration to a fully deployable one.}
\label{deploy}
\end{figure}

\noindent Our NIAC (NASA Innovative Advanced Concepts) phase I project with Joel Sercel on lunar-polar propellant mining outpost (LPMO): affordable exploration and industrialization, gives a deployable experimental model, shown in Figure~\ref{exper_model}. The metal materials for bars are available on the moon \cite{schrunk2007moon}, the strings (for example, UHMWPE, Ultra High Molecular Weight Polyethylene) can be shipped from the earth. There are mainly two ways to deploy tensegrity structures: 1. Altering string rest-lengths, which are usually realized by a motor-pulley-cable system \cite{abdallah2012position,sultan2014tensegrity}. 2. Using shape memory alloy (SMA) tendon wires, which are usually achieved by SMA and DC current supply devices \cite{bundhoo2009shape,kim2016soft}. The first method has these properties: wider control bandwidth, less cost, more environmentally robust, but mechanically more complicated than the second one. For this tower design, the deployability can be achieved by shorting the middle string length of the D-Bar. A shape control algorithm for class-k tensegrity \cite{chen2020design} can be applied to get the deploy sequence, shown in Figure \ref{deploy}.

% \begin{comment}

% \begin{figure}[H]
%     \centering
%     \includegraphics[width=0.3\textwidth]{t3d1_ground.png}
%     \caption{Three-dimensional $T_3D_1$ tensegrity tower.}
%     \label{t3d1_tower}
% \end{figure}

% \begin{figure}[H]
%     \centering
%     \includegraphics[width=0.3\textwidth]{model.PNG}
%     \caption{Three-dimensional $T_2D_1$ tensegrity tower.}
%     \label{t3d1_tower}
% \end{figure}
% \end{comment}

% \begin{comment}
% \begin{figure}[H]
%     \centering
%     \includegraphics[width=0.55\textwidth]{3d_tbar_q3.png}
%     \caption{Three-dimensional T-Bar system with complexity 3.}
%     \label{3d_tbar_system}
% \end{figure}
% \end{comment}

% \begin{comment}
% \begin{figure}[H]
%     \centering
%     \includegraphics[width=0.50\textwidth]{T3D1.png}
%     \caption{Three-dimensional $T_3D_1$ Bar structure.}
%     \label{3d_tdbar_system}
% \end{figure}
% \end{comment}

\section{Case Study}
~

\noindent The lunar craters vary in size from a few meters to 400 km, the depth of the craters are from less than 1 meter to 8 km, and a large portion of these craters is around 10 km in diameter and 2 km deep \cite{pike1974depth}. Here, we present a 2 km tall tower design, but the process developed in this paper is applicable for designing towers of any height. The payloads on the top of the tower include solar panels, communication devices, and mirrors with an estimated load of $m_p$ = 6,000 kg. Therefore, the compressive force at the top of the tower is $F = m_p g_{moon} = 6,000 \times 1.62$ N $= 9,720$ N. We use carbon-fiber rods for bars and UHMWPE for strings. The material properties are given in Table~\ref{tab:Material}. \par  
\begin{table}[H]
    \centering
     \caption{Material property for bars and strings, source: http://www.matweb.com/.}
    \begin{tabular}{llll}
    \toprule
Properties & Carbon-Rod & UHMWPE-String & Units \\ \midrule
Yield Stress & $1.72\times 10^9$ & $2.70\times 10^9$ &  Pa \\
Young's Modulus & $1.38\times 10^{11}$   & $1.20\times 10^{11}$ & Pa\\
Density & $1,500$  & $970$  & kg/m$^3$ \\
\bottomrule
    \end{tabular}
    \label{tab:Material}
\end{table}

~
\subsection{Mass of a Single Rod}
~

\noindent The mass required of a single carbon-fiber rod to take the compressive load $F$ subject to buckling constraints without considering gravity is: $m_{rod} = 2\rho_b H^2(\frac{F}{\pi E_b})^{\frac{1}{2}} = 2 \times 1500 \times 2000^2 \times  (\frac{9720}{3.14\times 1.38\times 10^{11}})^{\frac{1}{2}}~\text{kg} = 1.7973 \times 10^6 $ kg. Considering gravity, we put half of the gravity force $F_{rod} = \frac{1}{2}m_{rod}g_{moon}$ on the top of the rod (the compressive load becomes $F+F_{rod}$), then update the required mass $m_{rod}$ and $F_{rod}$, keep the iteration until the mass converges, we get $m_{rod,g} = 2.6919\times 10^8$ kg. \par  
~

\subsection{Mass of Tensegrity Structures}
~

\begin{figure}[H]
    \centering
    \includegraphics[width=0.8\textwidth]{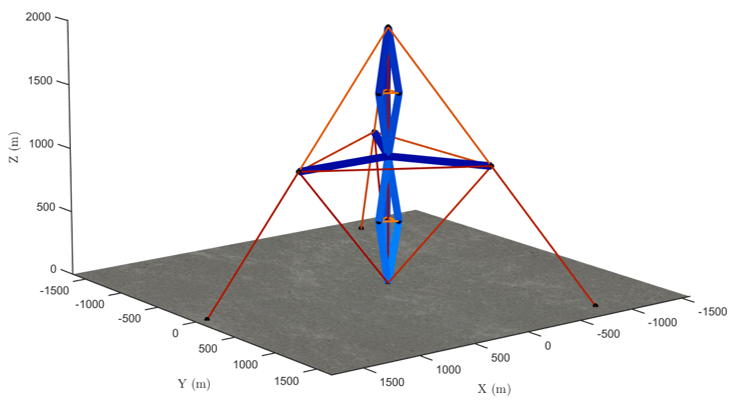}
    \caption{Three-dimensional $T_2D_1$ Bar structure with force $F$ applied at the top node and bottom nodes fixed. Maroon lines are strings and blue lines are bars, their thickness are scaled accordingly.}
    \label{3d_tdbar_system_final}
\end{figure}

\noindent To better illustrate the design idea, we reduce the variables of the design. That is, we propose the final configuration of the T-Bar and D-Bar angles of $T_qD_1$ tower to be $\alpha_T = \pi/3$ and $\alpha_D = \pi/18$. 
The optimization variables for the tower design are structure complexity, cross-sectional area of the members, and prestress in the strings. Algorithm \ref{alg1} was used in calculating the optimization variables to stabilize and minimize the total structure mass. The results are shown in Table \ref{tab:mass1}. Results show that $q=2$ with a lower bound of prestress in the strings to guarantees structure stability $\epsilon = 0.46$ is the optimal structure complexity with a total structure mass of $1.4081 \times 10^6$ kg. The detail information of the bars and strings are given in Table \ref{tab:mass2} and the structure configuration is shown in Figure \ref{3d_tdbar_system_final}.

\begin{table}[H]
    \centering \caption{Minimal mass of $T_qD_1$ tensegrity tower and prestress $\epsilon$, $\alpha_T = \pi/3$ and $\alpha_D = \pi/18$.}
    \begin{tabular}{lll}
    \toprule
Structure Complexity & Structure Mass & Prestress $\epsilon$ \\ \midrule
$q = 1$ & $1.7400 \times 10^6$ &  0.18 \\
$q = 2$ & $1.4081 \times 10^{6}$ & 0.46\\
$q = 3$ & $1.9683 \times 10^{6}$ & 2.10\\
\bottomrule
    \end{tabular}
    \label{tab:mass1}
\end{table}

\begin{table}[H]
    \centering \caption{Structure member information of the $T_2D_1$ tensegrity tower.}
    \begin{tabular}{lll}
    \toprule
Structure Member & Value & Units \\ \midrule
Bars Mass& $1.4081 \times 10^6$ &  kg \\
Strings Mass& $6.7733$ & kg\\
Total Mass& $1.4081 \times 10^6$ & kg\\
\bottomrule
    \end{tabular}
    \label{tab:mass2}
\end{table}

% \begin{table}[H]
%     \centering \caption{Minimal mass tensegrity tower, $\alpha_T = \pi/9$ and $\alpha_D = \pi/18$.}
%     \begin{tabular}{c|c|c}
%     \hline \hline
% Structure Complexity & Structure Mass & $\epsilon$ \\ \hline
% $q = 1$ & $1.9994 \times 10^6$ &  0.63 \\
% $q = 2$ & $2.3271 \times 10^{6}$ & 5.82\\
% \hline\hline
%     \end{tabular}
%     \label{tab:mass}
% \end{table}

% \begin{table}[H]
%     \centering \caption{Minimal mass tensegrity tower, $\alpha_T = \pi/18$ and $\alpha_D = \pi/18$.}
%     \begin{tabular}{c|c|c}
%     \hline \hline
% Structure Complexity & Structure Mass & $\epsilon$ \\ \hline
% $q = 1$ & $2.8183 \times 10^6$ &  2.46 \\
% \hline\hline
%     \end{tabular}
%     \label{tab:mass1}
% \end{table}

% \begin{table}[H]
%     \centering \caption{Minimal mass tensegrity tower, $\alpha_T = 2\pi/9$ and $\alpha_D = \pi/18$.}
%     \begin{tabular}{c|c|c}
%     \hline \hline
% Structure Complexity & Structure Mass & $\epsilon$ \\ \hline
% $q = 1$ & $1.7400 \times 10^6$ &  0.18 \\
% \hline\hline
%     \end{tabular}
%     \label{tab:mass1}
% \end{table}

~
\section*{CONCLUSIONS}
~

\noindent This paper presents a general framework to design minimal mass tensegrity towers. Starting from the compact matrix form of tensegrity statics, we first formulate structure mass subject to yielding and buckling of the structure members. Then, we model the gravitational force by equally distributing the total mass of the member on both end-nodes. The global stiffness matrix is also considered to guarantee the global stability of the structure. Then, we show the minimal mass problems can be formed as a non-linear programming problem, and an algorithm to solve this problem is also discussed. 
Finally, a case study is presented to design a deployable tensegrity tower. For a 2 km tall tensegrity tower made of carbon rods and UHMWPE strings to support 6000 kg payload in the presence of lunar gravity, $T_2D_1$ tower gives the minimal mass, $1.4081 \times 10^6$ kg, which is 0.52\% mass of a single 2 km rod $m_{rod,g} = 2.6919\times 10^8$ kg to support the same payload. The principles developed in this paper are also applicable to design many other static tensegrity structures. 

~
\section{Acknowledgement}
~ 

\noindent Some of the work in this paper was partially supported by the National Science Foundation under Award No. NSF CMMI-1634590 and by Texas A\&M University. The experimental model, shown in Figure \ref{exper_model}, was supported by a NIAC (NASA Innovative Advanced Concepts) phase I project, Lunar-Polar Propellant Mining Outpost (LPMO): Affordable Exploration and Industrialization, with partner Joel Sercel. The authors appreciate Mr. Ali Hasnain Khowaja for his help in building the experimental model.

%shown in Figure \ref{exper_model}. 

%This is a collaboration NIAC phase I project with Joel Sercel. 

% Lunar-Polar Propellant Mining Outpost (LPMO): Affordable Exploration and Industrialization

~

% add animation sequence
% write the deployable and mention working with Joel on the packing 

% akc: TransAstra Selected for Lunar Polar Mining Outpost NIAC Phase 1 Study 
% Lunar-Polar Propellant Mining Outpost (LPMO): Affordable Exploration and Industrialization

%\section*{ACKNOWLEDGEMENT}
% This work is funded by a NIAC phase II project, with partners Anthony Longman and Joel Sercel.
%\bibliographystyle{ascelike}
% \bibliography{ascexmpl}
\bibliography{ascexmpl}
\end{document}